\documentstyle[aps,psfig,amsmath]{revtex}
\newcommand{\be}{\begin{equation}}
\newcommand{\ee}{\end{equation}}
\newcommand{\bea}{\begin{eqnarray}}
\newcommand{\eea}{\end{eqnarray}}
\newcommand{\bef}{\begin{figure}}
\newcommand{\eef}{\end{figure}}
\newcommand{\bm}{\bibitem}
\newcommand{\al}{\alpha}
\newcommand{\bet}{\beta}
\newcommand{\gm}{\gamma}
\newcommand{\Gm}{\Gamma}
\newcommand{\et}{\eta}
\newcommand{\lm}{\lambda}
\newcommand{\Lm}{\Lambda}
\newcommand{\sg}{\sigma}
\newcommand{\Sg}{\Sigma}
\newcommand{\de}{\delta}
\newcommand{\De}{\Delta}
\newcommand{\Dell}{\Delta_{11}}
\newcommand{\gf}{\gamma_5}
\newcommand{\ep}{\epsilon}

\newcommand{\om}{\omega}
\newcommand{\rw}{\rightarrow}

\newcommand{\mn}{\mu\nu}
\newcommand{\cl}{\cal{L}}

\newcommand{\bt}{{\boldsymbol{\tau}}}
\newcommand{\bde}{{\boldsymbol{\Delta}}}
\newcommand{\bsg}{{\boldsymbol{\Sigma}}}
\newcommand{\bgm}{{\boldsymbol{\Gamma}}}

\newcommand{\U}{{\boldsymbol{U}}}

\newcommand{\T}{{\boldsymbol{T}}}
\newcommand{\F}{F_\pi}
\newcommand{\M}{M_\pi}

\newcommand{\del}{\partial}
\newcommand{\dmd}{\partial_\mu}
\newcommand{\dmu}{\partial^\mu}
\newcommand{\dnd}{\partial_\nu}

\newcommand{\p}{\vec{\phi}}    
\newcommand{\vp}{\vec{p}} 

\newcommand{\vq}{\vec{q}}
\newcommand{\la}{\langle}
\newcommand{\ra}{\rangle}

\newcommand{\pr}{^{\,\prime}}

\newcommand{\ta}{\tau^a}
\newcommand{\ola}{\overline{l}_1}
\newcommand{\olb}{\overline{l}_2}
\newcommand{\olc}{\overline{l}_3}
\newcommand{\old}{\overline{l}_4}
\newcommand{\oli}{\overline{l}_i}
\newcommand{\oj}{\overline{J}}
\newcommand{\ok}{\overline{K}}
\newcommand{\ot}{\overline{T}}

\begin{document}

\setcounter{page}{1}

\title{Pion propagation in real time field theory at finite temperature}

\author{S. Mallik} 
\address{Saha Institute of Nuclear Physics,
1/AF, Bidhannagar, Kolkata-700064, India} 

\author{Sourav Sarkar} 
\address{Variable Energy Cyclotron Centre, 1/AF, Bidhannagar, Kolkata-700064,
 India}

%\date{} 

\maketitle

\begin{abstract} 

We describe how the thermal counterpart of a vacuum two-point function may
be obtained in the real time formalism in a simple way by using directly the 
$2\times 2$ matrices that different elements acquire in this formalism. Using this
procedure we calculate the analytic (single component) thermal amplitude for 
the pion pole term in the ensemble average of two axial-vector currents to two 
loops in chiral perturbation theory. The general expressions obtained for the
effective mass and decay constants of the pion are evaluated in the chiral and
the nonrelativistic limits. We also investigate the effect of massive states 
on these effective parameters.
 
\end{abstract}

%\pacs{PACS numbers: 11.10.Wx, 12.38.Mh, 12.39 Fe}

\section{Introduction}

The real time thermal field theory is apparently complicated by the fact 
that all two-point functions in this formalism assume the form of 
$2\times 2$
matrices \cite{Umezawa}. These matrices, however, have simple structures: 
If we factor out certain matrices depending on the distribution function
only, these become diagonal, each with essentially a single independent
element with proper analytic properties. But in actual computations one
tends to ignore the matrix structure, starting instead with the so-called
physical $11$-element, encountering though summation over indices at all
interaction vertices in a Feynman graph. Such a procedure encounters
ill-defined products of components of the matrix propagator, that must be
combined together to get a well-defined quantity. Further the 11-component
does not have a simple analytic structure.  

In this work we show that it is both simple and elegant to work with the
matrix amplitudes. All one has to do is to write out the usual vacuum
amplitude. The thermal matrix amplitude is then obtained by replacing its 
elements like the propagator, the self-energy and the vertices by the 
corresponding matrices. Factorizing these matrices as mentioned above, we 
immediately get the analytic amplitude representing the dynamics of the 
system in the heat bath.

Here we apply this procedure to calculate the pion pole term in the 
two-point function of the axial-vector currents to two loops in chiral
perturbation theory \cite{Gasser}. This problem was studied earlier by
several authors \cite{Leutwyler1,Schenk,Song}, in particular, by Toublan 
\cite{Toublan}. After obtaining the analytic amplitude, we follow him to find 
the pion pole position and the residue. We then find these pole parameters in 
the chiral limit, in agreement with his results. We also evaluate them in the 
nonrelativistic region and consider the effect of massive states on them.

In Sec. II we write down the effective chiral Lagrangian to fourth order,
needed to obtain all the required vertices.
In the next Sec. III, we obtain the vacuum amplitude from all the Feynman
graphs up to two loops contributing to the pion pole term of the two-point
function. The corresponding thermal amplitude is obtained in Sec.IV, from
which we derive the effective parameters, namely the pion mass and the decay
constants at finite temperature. These expressions are evaluated 
analytically in Sec. V in the high and low temperature limits. We examine
the contribution of the massive states in Sec. VI. Finally we bring out the main 
features of our work in Sec. VII.

Appendix A constitutes an essential part of this work. Reviewing briefly the
real time thermal field theory, we discuss here at length how the vacuum
amplitude for an individual Feynman graph may be converted into its thermal
counterpart. In Appendix B we write the integrals appearing in the
nonfactorizable amplitudes. In the last Appendix C we collect the results for
the relevant integrals in the high and the low temperature region.

\section{Chiral Perturbation Theory} 
\setcounter{equation}{0}
\renewcommand{\theequation}{2.\arabic{equation}}

We consider the $QCD$ Lagrangian for the doublet of light quarks, $u$ and
$d$. In the absence of their masses, it has chiral symmetry, being invariant
under $SU(2)_R\times SU(2)_L$. This symmetry is supposed to be broken 
spontaneously to $SU(2)_V$ of ordinary isospin, generating the massless
pions as the Goldstone bosons. 

In the physical case of non-zero quark masses, chiral symmetry is also
broken explicitly to the same isospin subgroup, if we neglect the mass
difference of $u$ and $d$ quarks. The pions become the pseudo-Goldstone
bosons acquiring mass $M$, given by $M^2=2m_qB$ to lowest order, where $B$
is related to the quark condensate in vacuum, whose dynamical generation
leads to the spontaneous symmetry breaking.
 
As already stated, we are interested in the pion pole term in 
the two point function of the axial-vector currents,
\[A_\mu^a=\overline{q}\gm_\mu\gf\frac{\ta}{^2}q\,~~~~~~a=1, 2, 3\,,\]
evaluated in chiral perturbation theory, the effective theory of $QCD$ at low
energy. Such functions are best calculated in the external field method, in 
which one introduces in the original $QCD$ Lagrangian an external field 
$a^\mu_a(x)$ coupled to $A_\mu^a(x)$ as well as a field $v^\mu_a(x)$ for the 
vector currents \cite{Gasser}. The global chiral symmetry is then promoted 
to a local one, with appropriate transformation properties of the external 
fields.

In the effective theory, the pion fields may be collected in the form of
an unitary matrix,
\[U(x)=e^{i\phi^a(x)\ta/F}~,\]
where $\ta$ are the Pauli matrices. The constant $F$ may be identified as
the pion decay constant in the chiral limit. The local symmetry requires us to 
replace the ordinary derivative by the covariant one,
\be
D_\mu U=\partial _\mu U-i\{a_\mu,U\}\,, 
\ee
where, for our purpose, we retain only the external field $a_\mu (x)$. 
The two-point function of $A_\mu (x)$ is now obtained as the coefficient of the
quadratic term in $a_\mu (x)$ in the perturbative evaluation of the
generating functional with the effective Lagrangian.

As a non-renormalizable theory, the effective Lagrangian consists of a series 
of terms with an increasing number of derivatives and/or quark mass factors,
\[{\cl}_{e\!f\!f}={\cl}^{(2)}+{\cl}^{(4)}+\cdot\cdot\cdot\cdot~.\]
The leading term is given by
\be
{\cl}^{(2)}=\frac{F^2}{4}\{\la D_\mu U^\dag D^\mu U\ra 
+ M^2\la U+U^\dag\ra\}\,,
\ee
where $\la A\ra$ denotes the trace of the 2$\times$2 matrix $A$. 
Here the first term is invariant under the chiral transformations. The second 
term represents explicit symmetry breaking due to the quark mass term 
in the $QCD$ Lagrangian.
 
The next, non-leading piece in ${\cl}_{e\!f\!f}$ is \cite{Gasser,Gerber} 
\be
{\cl}^{(4)}=\frac{1}{4}l_1 \la D_\mu U^\dagger D^\mu U\ra^2+
\frac{1}{4}l_2 \la D_\mu U^\dagger D_\nu U\ra\la D^\mu U^\dagger D^\nu U\ra
+\frac{1}{8}l_4 M^2\la D_\mu U^\dagger D^\mu U\ra\la U+U^\dagger\ra
+\frac{1}{16}(l_3+l_4) M^4\la U+U^\dagger\ra^2~.
\ee
It provides counterterms necessary to renormalize the one-loop graphs with
vertices from ${\cl}^{(2)}$. Thus 
the bare coupling constants $l_1,\cdots,l_4$ contain a pole at $d=4$ in
dimensional regularization. The coefficients of these poles may be
determined by evaluating all the one-loop graphs. Alternatively, these
may be obtained directly by calculating the
short distance behaviour of the generating functional to one loop
\cite{Gasser}. Adopting the notation introduced in this Ref., 
the renormalized coupling constants $\ola, \cdots, \old$  are defined by,
\be
l_i=\gm_i \left(\lm +\frac{1}{32\pi^2}\oli\right),
\ee
with
\[\gm_1=\frac{1}{3},~~~ \gm_2=\frac{2}{3},~~~ \gm_3=-\frac{1}{2},~~~
\gm_4=2\,. \]
The pole is contained in $\lm$,
\be
\lm=\frac{M^{d-4}}{(4\pi)^2} \left( \frac{1}{d-4}-\frac{1}{2}[\ln 4\pi 
+\Gm'(1)+1]+O(d-4)\right)\,.
\ee
Up to the factor $\gm_i/32\pi^2$, the constants $\oli$ are running coupling
constants at the scale $M$. The $M$-dependence of $\oli$ can be made
explicit by relating these to $l_i^r$, the renormalized coupling constants at
any other scale $\mu$,
\be
l_i^r =\frac{\gm_i}{32\pi^2}\left(\oli +\ln\frac{M^2}{\mu^2}\right)\,. 
\ee
In Sec.V we shall use this equation to show the finiteness of the pion pole
parameters in the chiral limit.
  
The renormalization of the two-loop graphs would require vertices from the
next higher piece, $ {\cl}^{(6)}$ in  ${\cl}_{e\!f\!f}$. But we do not need
it, as we are not interested in the vacuum amplitudes, but only in their
temperature dependent parts.

\section{Vacuum Amplitude}
\setcounter{equation}{0}
\renewcommand{\theequation}{3.\arabic{equation}}

Here we obtain the pion pole contribution to the vacuum two-point
function of the axial-vector current, 
\be
\de^{ab} T_{\mn}(q)=i\int d^4x\, e^{iq\cdot x}\,\la 0 |\,T\,A_\mu^a(x)
\,A_\nu^b(0)\,| 0 \ra_{\rm{\pi\, pole}}\,,
\ee
by calculating all the Feynman graphs up to two loops with the interaction
vertices given by ${\cl}^{(2)}$ and ${\cl}^{(4)}$. These graphs are conveniently
divided into four groups, shown in Figs. 1-4.

\bef
\centerline{\psfig{figure=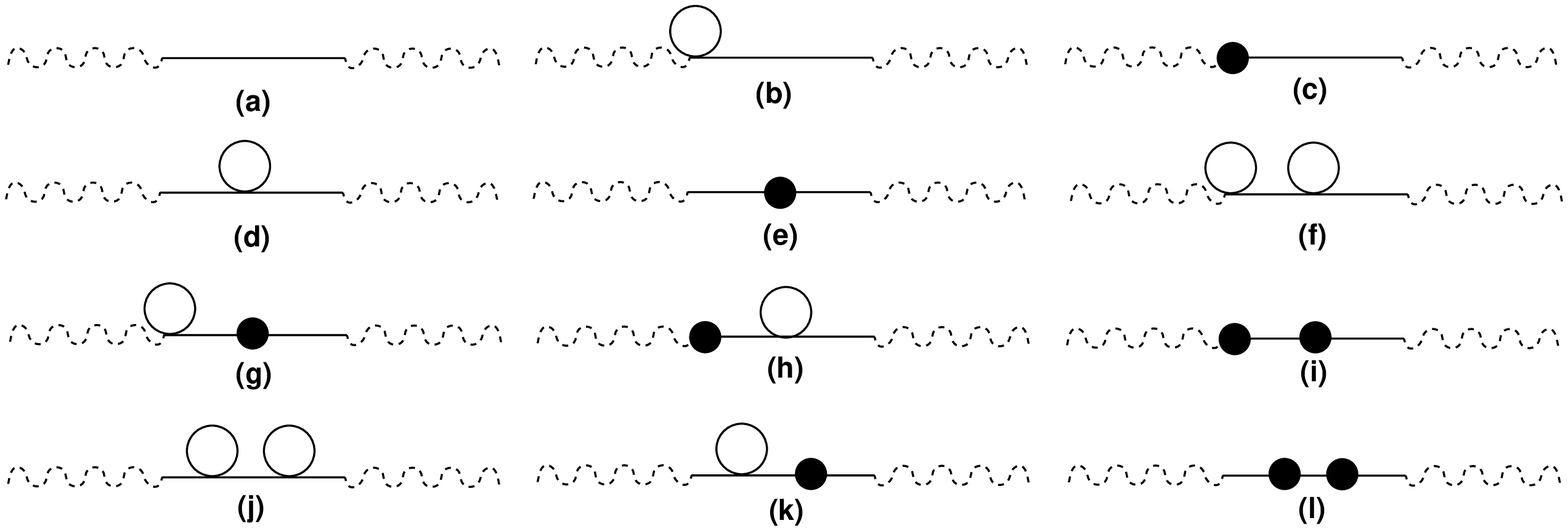,height=3.8cm,width=11cm}}
\caption{The free amplitude with corrections from one-loop graphs along with
counterterm graphs and those two-loop graphs that are iterations of the
former ones. Vertices of ${\cl}^{(2)}$ and ${\cl}^{(4)}$ are shown as points
and filled circles respectively. Wavy and straight lines denote axial
current and pion respectively.} 
\eef

\bef
\centerline{\psfig{figure=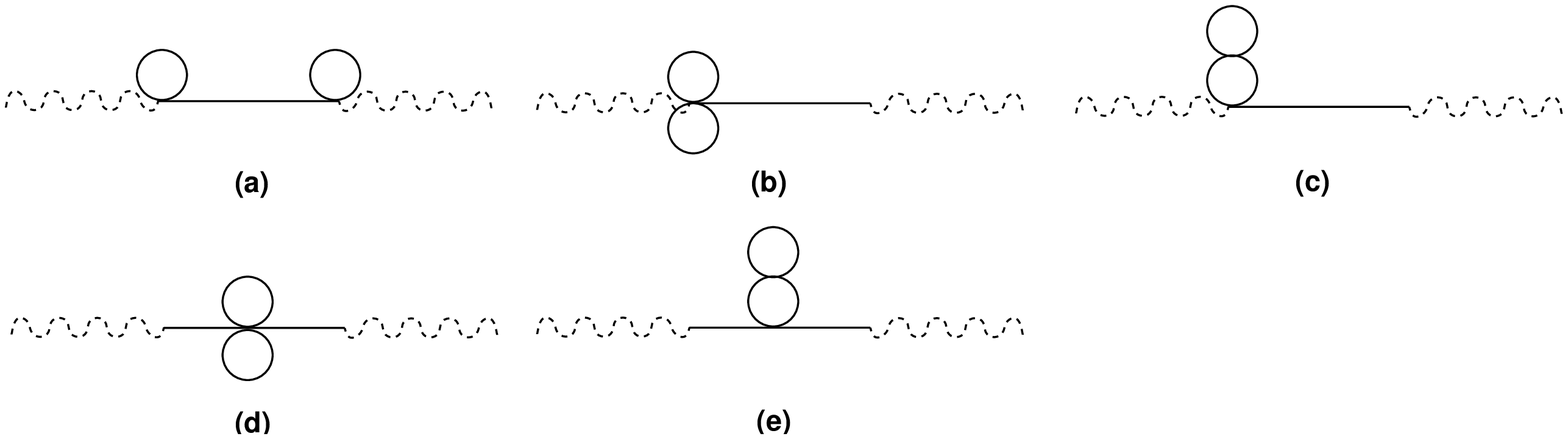,height=3cm,width=11cm}}
\caption{Factorizable two-loop graphs with vertices from ${\cl}^{(2)}$ only}
\eef

\bef
\centerline{\psfig{figure=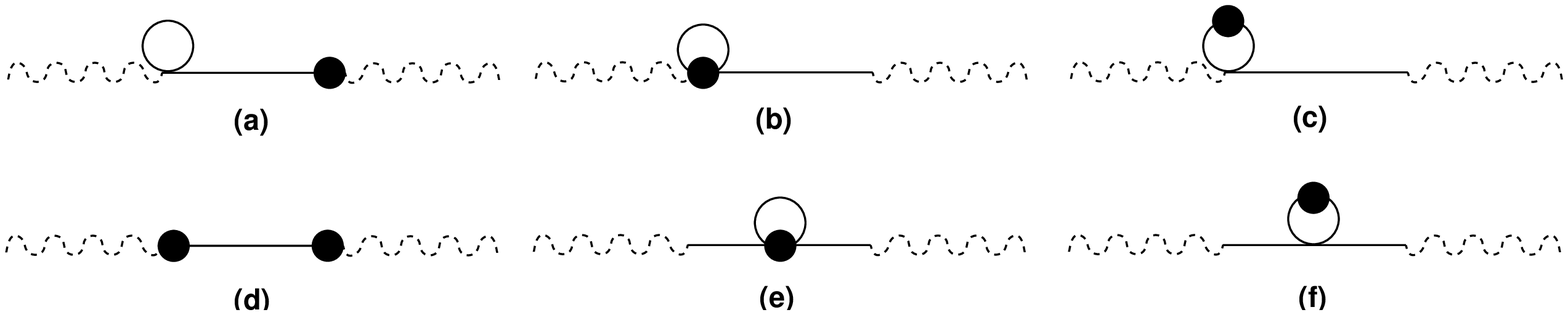,height=2.25cm,width=11cm}}
\caption{Further counterterm graphs}
\eef

\bef
\centerline{\psfig{figure=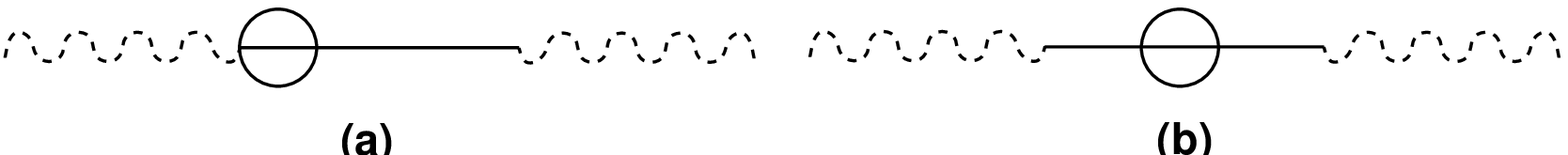,height=0.7cm,width=7.5cm}}
\caption{Nonfactorizable two-loop graphs}
\eef

Let us first derive the familiar one loop results. 
The vertex correction graphs (b) and (c) of Fig.~1 modify the residue of
the free amplitude of graph (a) to give 
\be
T^{(1a,b,c)}_{\mn}(q)=
q_\mu q_\nu F^2\left\{1+4\et (3l_4-2J)/3\right\}i\De(q)~~.
\ee
where $\et=\displaystyle\frac{M^2}{F^2}$ is an expansion parameter and
$\De (q)$ is the free pion propagator, $\De (q)=i/(q^2 -M^2 +i\ep)$. $J$ 
is a divergent one-loop integral,
\be
J(M)=\frac{1}{M^2}\int\frac{d^4k}{(2\pi)^4}\,\De(k)\equiv 2\lm \,,
\ee
with $\lm$ given by Eq.(2.5).
To include the self-energy graphs, it is convenient to introduce here the
well-known Dyson-Schwinger equation for the complete propagator 
$\De\pr (q)$,
\be
\De\pr(q)=\De(q)+\De(q)(-i\Sg(q))\De\pr(q)
\ee
where the self-energy part $\Sg$ of graphs (d) and (e) of Fig.~1 is given 
by
\be
\Sg(q)=-2\et(3l_4-J)(q^2-M^2)/3+F^2\et^2(4l_3+J)/2~.
\ee
Eq.(3.4) may be solved by iteration,
\be
\De\pr(q)=\De(q)+\De(q)(-i\Sg(q))\De(q)+\De(q)\{-i\Sg(q)\De(q)\}^2+\cdots
\ee
or in closed form,
\be
\De\pr(q)=\frac{\De(q)}{1+i\Sg(q)\De(q)}~~.
\ee
The self-energy correction is now included in Eq.(3.2) by replacing $\De(q)$ 
with $\De\pr(q)$. Thus we get the one-loop result for the pion pole,
\[T_{\mn}(q)=-q_\mu q_\nu \frac{\F^2}{q^2-M_\pi^2+i\ep} ,\]
with
\be
M_\pi^2=M^2\left\{1+2\et(l_3+J/4)\right\}=M^2(1-\et \olc/32\pi^2)\,,
\ee
\be
\F=F\left\{1+\et(l_4-J)\right\}= F(1+\et \old/16\pi^2)\,,
\ee
on using Eq.(2.4).

We now include the two-loop graphs. First consider those of Figs. 1, 2 and 3,
that are actually products of two one-loop parts. The total contribution of
all these graphs may be put in the form
\be
T^{(1+2+3)}_{\mn}(q)=q_\mu q_\nu F^2\sum_{n=1}^3\{\gm_n\,i\De(q)+
\sg_n M^2 \De^2(q)\}
+8F^2\et^2(l_1+2l_2)(q_\mu J_{\nu\lm}+q_\nu J_{\mu\lm})q^\lm
\,i\De(q)~,
\ee
where we show separately the sums of contributions of all graphs in each of
Figs.1, 2 and 3. Thus the sum of graphs in Fig.1 is given by the $n=1$ term
\footnote{A piece, namely $-q_\mu q_\nu (F^2/4)\eta^2
(J+4l_3)^2M^4i\De^3(q)$, is omitted here, as it is automatically included
when we put the $n=1$ term in the form of a simple pole.}. Note here that the 
two loop graphs from (f) to (l), being
iterations of graphs from (b) to (e), are automatically included in
Eq.(3.7). But we prefer to write them explicitly in Eq.(3.10),
getting 
\bea
\gm_1&=&1+2\et(l_4-J)-4\et^2(l_4-J)(3l_4-J)/3~,
\nonumber\\
\sg_1&=&\et (4l_3+J)/2-2\et^2 J(4l_3+J)/3~,
\eea
where we cancel factors as, $(q^2-m^2)\De^2(q)=i\De(q)$, that can also be
justified at finite temperatures. In this way we get directly the shifts in
the residue and the pole position from graphs of Fig.1 as $F^2(\gm_1 -1)$
and $M^2\sg_1/\gm_1$ respectively. Next, the graphs of Fig.~2 contain only 
the vertices of ${\cl}^{(2)}$ and their sum is given by the $n=2$ term, with
\be
\gm_2=\et^2J(8J+3J\pr)/3~,~~~~~\sg_2=-\et^2 J(3J+2J\pr)/8
\ee
where we encounter a new divergent integral related to the earlier one as, 
\be
J\pr(M)=i\int\frac{d^4k}{(2\pi)^4}\,\De^2(k)=-\frac{\partial}
{\partial M^2}(M^2J)=-2\lm-\frac{1}{16\pi^2}\,.
\ee
Lastly, the sum of  graphs of Fig.~3 with vertices from ${\cl}^{(2)}$ and
${\cl}^{(4)}$ is given by the term $n=3$ with
\bea
\gm_3&=&\et^2\{(36l_1+12l_2-25l_4)J+12l_3J\pr+12l_4^2\}/3\,,\nonumber\\
\sg_3&=&-\et^2\{(36l_1+12l_2+16l_3-3l_4)J+3l_3 J\pr
+24(l_1+2l_2)q^\lm q^\sg J_{\lm\sg}/M^2\}/3\,,
\eea
together with the remaining term in Eq.(3.10), where we have still another
divergent integral,
\be
J_{\mn}(M)=\frac{1}{M^4}\int\frac{d^4k}{(2\pi)^4}\,k_\mu k_\nu\De(k)\,.
\ee
Actually this term is also proportional to $q_\mu q_\nu$, once the integral
is evaluated. But in view of its extension to finite temperature, we keep 
it as such.

Finally we have the amplitude from the nonfactorizable two-loop graphs of 
Fig.~4, 
\be
T_{\mn}^{(4)}(q)=-\frac{2i}{9F^2}\left\{q_\mu \De(q)\Gm_\nu(q)+\Gm_\mu(q)
\De(q)q_\nu\right\}+ \frac{1}{18F^2}q_\mu q_\nu \De(q)\Sg(q)\De(q)~,
\ee
where the vertex function $\Gm_\mu(q)$ of graph (a) is 
\be
\Gm_\mu(q)=i\int\frac{d^4k_1}{(2\pi)^4}\frac{d^4k_2}{(2\pi)^4}
(2q_\mu-3k_{1\mu}-3k_{2\mu})f\De(k_1)\De(k_2)\De(q-k_1-k_2)~,
\ee
and the self-energy function $\Sg(q)$ of graph (b) is
\be
\Sg(q)=-i\int\frac{d^4k_1}{(2\pi)^4}\frac{d^4k_2}{(2\pi)^4}(3M^4+2f^2)
\De(k_1)\De(k_2)\De(q-k_1-k_2)
\ee
with $f$ standing for the function,
\[f(q,k_1,k_2)=k_1^2+k_2^2+4k_1k_2+M^2+2q\cdot(k_1+k_2)-2q^2 \,.\]
The order of the factors in Eq.(3.16) is in 
anticipation of their matrix structures of the thermal amplitudes in the 
next section.

In Ref.~\cite{Toublan} the vertex and self-energy integrals
have been cast in a particularly convenient form using the symmetries of the
integrands under the interchange of the integration variables. Thus if one 
defines
\be
K(q)=\frac{i}{M^2}\int\frac{d^4k_1}{(2\pi)^4}\frac{d^4k_2}{(2\pi)^4}
\De(k_1)\De(k_2)\De(q-k_1-k_2)
\ee
\be
K_{\mn}(q)=\frac{i}{M^4}\int\frac{d^4k_1}{(2\pi)^4}\frac{d^4k_2}{(2\pi)^4}
k_{1\mu}k_{1\nu}\De(k_1)\De(k_2)\De(q-k_1-k_2)~,
\ee
they may be written as
\be
\Gm_\mu(q)=-2F^4\et^2[q_\mu(3J^2-K)+9K_{\mu\rho}q^\rho]\,,
\ee
\be
\Sg(q)=-F^4\et^2[4(9J^2-4K)(q^2-M^2)+3M^2(8J^2-K)+72q^\rho q^\sg K_{\rho\sg}]\,.
\ee
Then Eq.(3.16) simplifies to
\be
T^{(4)}_{\mn}=q_\mu q_\nu F^2\et^2\left\{\frac{2}{3}J^2
i\De(q)-\frac{1}{6}(8J^2-K
+24q^\rho q^\sg K_{\rho\sg}/M^2)M^2\De^2(q)\right\}
+4F^2\et^2(q_\mu K_{\nu\lm}+q_\nu K_{\mu\lm})q^\lm
\,i\De(q)\,.
\ee

The sum of amplitudes (3.10) and (3.23), along with the one from the tree
graphs with a single insertion of vertices from $ {\cl}^{(6)}$ (not 
calculated above)  would give the complete, renormalized vacuum amplitude.
One may then extend the one-loop results (3.8, 3.9) for the pion pole
parameters to two-loops. Instead, however, we turn to the corresponding 
thermal amplitudes to find the temperature dependence of these parameters.
 
\section{Thermal Amplitude}
\setcounter{equation}{0}
\renewcommand{\theequation}{4.\arabic{equation}}

The thermal (ensemble averaged) two-point function of the axial-vector current is  
a $2\times 2$ matrix in the real time formalism, whose $ij$-th element is 
\be
(T^{ab}_{\mn})_{ij}=i\int d^4x e^{iq\cdot x} {\rm Tr} [\rho T_c A^a_\mu(\phi_i(x))\,
A^b_\nu(\phi_j(x))]/{\rm Tr} \rho\, |_{\rm{\pi~ pole}},~~~\rho = e^{-\beta H},
\ee
where $T_c$ denotes 
time ordering with respect to the time contour of Fig.9 in Appendix A. There we
discuss at length how to obtain the matrix amplitude for an
individual graph from its vacuum amplitude. To summarize,
all we need is to replace the loop integrals $(J,\,J',\,J_{\mn})$
encountered in the vacuum amplitudes by $(J^\bet,\,J^{\prime\,\bet},\,
J^\bet_{\mn})$, where, in effect, the vacuum pion propagator is replaced by
the $11$- or $22$- component of the thermal propagator. Further, the elements
of the vacuum theory, namely, $(\De,\,\Sg,\,\Gm)$ need be replaced by the
matrices $(\bde,\,\bsg,\,\bgm)$ and also $\De^2$ by $\bde\bt\bde$, where the
matrices are given in Appendix A.
Having obtained the matrix amplitude, we put in the factorized forms for all
the matrices to get an equation among diagonal matrices, each of
whose $11$- and $22$- elements are identical up to complex
conjugation and possibly a $(-)$ sign. Thus we leave behind the matrix 
structure and work with the (single component) analytic amplitude 
$T_{\mn}^\bet (q)$. Almost repeating Eqs.~(3.10) and (3.23) we write it as
the sum of
\be
T^{\bet\, (1+2+3)}_{\mn}=q_\mu q_\nu F^2\sum_{n=1}^3\{\gm_n^\bet i\De(q)
+\sg_n^\bet M^2 \De^2(q)\}
+8F^2\et^2(l_1+2l_2)(q_\mu J^\bet_{\nu\lm}+q_\nu J^\bet_{\mu\lm})q^\lm
\,i\De(q)~,
\ee
where $\gm_n^\bet$ and $\sg_n^\bet$ are obtained from $\gm_n$ and $\sg_n$
of Eqs.(3.11), (3.12) and (3.14) after replacing $J$'s by $J^\bet$'s, and 
\be
T^{\bet\,(4)}_{\mn}=q_\mu q_\nu F^2\et^2\left[\frac{2}{3}(J^\bet)^2
i\De(q)-\frac{1}{6}\{8(J^\bet)^2-K^\bet
+24q^\rho q^\sg K^\bet_{\rho\sg}/M^2\}M^2\De^2(q)\right]
+4F^2\et^2(q_\mu K^\bet_{\nu\lm}+q_\nu K^\bet_{\mu\lm})q^\lm
\,i\De(q)
\ee
where, as in Eqs.(3.21) and (3.22) for the vacuum case, we have expressed 
$\Gm^{\bet}$ and $\Sg^{\bet}$ in terms of $ K^\bet,\,K_{\mn}^\bet$  
(and $J^\bet$) with
\be
K^\bet (q)=\frac{i}{M^2}\int\frac{d^4k_1}{(2\pi)^4}\frac{d^4k_2}{(2\pi)^4}
\Dell(k_1)\Dell(k_2)\Dell(q-k_1-k_2)
\ee
and
\be
K^\bet_{\mn}(q)=\frac{i}{M^4}\int\frac{d^4k_1}{(2\pi)^4}\frac{d^4k_2}{(2\pi)^4}
k_{1\mu}k_{1\nu}\Dell(k_1)\Dell(k_2)\Dell(q-k_1-k_2)~.
\ee
We must point out here that Eqs.(4.4) and (4.5) hold only for the {\it real}
parts. (The {\it imaginary} parts of both sides differ by the factor
$(1+2n(|q_0|))^{-1}$, as follows from Eqs.(A.27) and (A.31).) Since,
however, we are interested only in the real parts of the pole parameters, our
imprecise notation will lead to no error.

As already stated, the vacuum part of this thermal amplitude
is of no interest to us here, beyond the one loop results given in Eqs.(3.8)
and (3.9). In the $\bet$-dependent part, we have to isolate the finite terms
from the divergent ones. To this end, we separate the $\bde_{11\, or 22}$
into its vacuum and thermal parts in the expressions for $J^\bet$'s and 
$K^\bet$'s. In the case of $J^\bet$'s we have simply,  
\[J^\bet=J+\oj~,~~~J^{\prime\,\bet}=J\pr+\oj\pr~,~~~
J^\bet_{\mn}=J_{\mn}+\oj_{\mn}\,,\]
where
\be
(\oj,\,\oj\pr,\,\oj_{\mn} )=\int\frac{d^4k}{(2\pi)^3}n(|k_0|)
\left (\frac{1}{M^2},-\frac{\partial}{\partial M^2},
\frac{k_\mu k_\nu}{M^4}\right ) \de(k^2-M^2)\,.
\ee
$K^\bet$ splits as
\be
K^{\bet}(q)=K(q)+K^{\bet}(q)|_{div}+\ok^J+\ok (q)\,.
\ee
Here the second term gives the $\bet$-dependent divergent pieces, all 
proportional to $\lm$. The finite, temperature dependent part can be expressed 
partly in terms of $\oj$'s defined above and the remainder as certain 
$q$-dependent integrals, constituting the third and the fourth term
respectively. A similar decomposition holds for $K_{\mn}^\bet$,
\be 
K_{\mn}^{\bet}(q)=K_{\mn}(q)+K_{\mn}^{\bet}(q)|_{div}+\ok_{\mn}^J
+\ok_{\mn}(q)\,.
\ee
All the pieces in Eqs.(4.7) and (4.8) are displayed in Appendix B. 

It is now easy to see that all the $\bet$-dependent divergent pieces cancel
out. We then get the complete, renormalized thermal amplitude as,
\be
T^{\bet}_{\mn}(q)= T\pr _{\mn} (q) +\ot_{\mn} (q)\,,
\ee
where $T\pr_{\mn}$ is the vacuum amplitude without the free pole term,
which is put in the $\bet$-dependent piece $\ot_{\mn}$ 
\bea
\ot_{\mn}(q)=& & q_\mu q_\nu F^2\left[ (1-2\oj\eta +A\eta^2)i\De(q)+
\left\{ \frac{1}{2} \oj\eta -\left( B-\frac{\ok (q)}{6} +\frac{q^\lm q^\sg}
{M^2} S_{\lm \sg} (q)\right) \eta^2 \right\} M^2\De^2(q) \right]\nonumber \\
& & + F^2 \{q_\mu S_{\nu\lm}(q) +q_\nu S_{\mu\lm}(q)\}q^\lm \eta^2 i\De(q)\,,
\eea
where the tensor $S_{\mn}$ is given by
\be
S_{\mn}(q)=l\,\oj_{\mn}+4\ok_{\mn}(q)~,
\ee
and the ($\bet$-dependent) constants $A$ and $B$ are built out of $\oj$
and $\oj\pr$,
\be
A=\oj(2\oj+l\pr)+\oj\pr\left(\oj-\frac{\olc}{16\pi^2}\right)\,,~~~~
B=\oj(\frac{19}{8}\oj+l^{''})+\frac{\oj\pr}{4}\left(\oj-
\frac{\olc}{16\pi^2}\right)\,.
\ee
The three combinations of coupling constants introduced above are,
\be
l=\frac{1}{12\pi^2}\left(\ola+4\olb-\frac{14}{3}\right)\,,~~
l\pr=\frac{1}{48\pi^2}\left(6\ola+4\olb-9\old-\frac{7}{3}\right)\,,~~
l^{''}=\frac{1}{48\pi^2}\left(6\ola+4\olb-6\olc-3\old-\frac{55}{12}\right)~.
\ee

In Eq.(4.10) we have left out regular, non-pole pieces arising out of 
explicit factors of $q^2$. We remove further non-pole pieces by expanding 
the $q$-dependent functions in $q_0$ in the neighbourhood of the pole 
$q_0^2=\vq\,^2+M^2\equiv \om^2$ at fixed $\vq$. Thus
\be
\ok (q)=\ok^{(0)} (\om)+ (q_0^2-\om^2) \ok^{(1)} (\om)+\cdots\,,~~~~~ 
\ok^{(1)} (\om)=\frac{1}{2\om }\frac{\partial}{\partial q_0}\ok (q)
|_{q_0 =\om}\,,
\ee
and similarly for $S_{\mn} (q)$. The resulting expression may be put in 
the form
\be
\ot_{\mn} (q)=-\frac{f_\mu (q) f_\nu (q)}{q_0^2-\Omega^2 (\om)}\,,
\ee 
where
\bea
\Omega^2(\om)& &= \vq\,^2+M^2\left\{1+\frac{1}{2}\oj\eta
-\left( B-\frac{\ok^{(0)}}{6}+\frac{1}{M^2}(q^\lm q^\sg
S_{\lm\sg})^{(0)}-\oj^2\right)\eta^2\right\}\nonumber \\
f_\mu (q)& &=F\left[ q_\mu\left\{ 1-\oj\eta
+\frac{1}{2}\left( A+\frac{M^2}{6}\ok^{(1)}-(q^\lm q^\sg
S_{\lm\sg})^{(1)}-\oj^2\right)\eta^2\right\}+(S_{\mu\lm}q^\lm)^{(0)} 
\eta^2\right]\,.
\eea
Note that $f_\mu(q)$ is not proportional to $q_\mu$ due to presence of
the last term. This is due to the lack of Lorentz invariance in a medium,
which serves as the preferred frame of reference. Thus, as in nonrelativistic 
systems \cite{Leutwyler2}, we have here two different $F$'s, the temporal and 
the spatial \cite{Pisarski},
\be
f_0(q)=q_0F^t(q), ~~~~~~~~~f_i(q)=q_iF^s(q)
\ee

One may now find the thermal dispersion curve for the pion and its decay 
'constants' at different values of $|\vq|$. Instead, however, we set $\vq =0$ 
and find the effective mass and the decay constants as a function of 
temperature. Converting the parameters $F$ and $M$ to the physical values
by Eqs.(3.8) and (3.9) and using the result (B.9) of Appendix B, we 
finally get them as
\bea
&&\M^2(T)=\M^2\left\{1+\frac{\M^2}{2\F^2}\oj - \frac{\M^4}{\F^4}\left(
l^{'''}\oj+\frac{11}{8}\oj^2+\frac{1}{4}\oj\,\oj^{'}-\frac{\ok}{6}+
l\, \oj_{00}+4\ok_{00}\right)\right\}\,,\\
&&\F^t=\F\left\{1-\frac{\M^2}{\F^2}\oj+\frac{\M^4}{2\F^4}\left(
\oj(\oj+l^{'})+ \oj\,\oj^{'}+\frac{\M}{12}\frac{\del\ok}{\del q_0}+
l\,\oj_{00}+4\ok_{00}-2\M\frac{\del\ok_{00}}{\del q_0}\right)\right\}\,,\\
&&(\F^t(T)-\F^s(T))/\F=\frac{\M^4}{3\F^4}\left\{-12\M C-l\,\oj+
4(\oj^2-\ok+l\,\oj_{00}+4\ok_{00})\right\}\,,
\eea
where $l^{'''}$ is given by
\be 
l^{'''}=\frac{1}{192\pi^2}\left(24\ola+16\olb-27\olc-24\old
-\frac{55}{3}\right)\,,
\ee 
and $C$ is the coefficient of the linear 
term in the expansion of $\ok_{0i}(q)$ around $\vq = 0$,
\be
\ok_{0i}(q)=C(q_0)q_i + \cdots .
\ee
Note that all the quantities $\oj$, $\ok$ etc. are now functions of $\M$. 
These results agree with Ref.\cite{Toublan} except for the definition of 
$l^{'''}$ \cite{Comment}.

\section{Evaluation}
\setcounter{equation}{0}
\renewcommand{\theequation}{5.\arabic{equation}}

We now need the values of the coupling constants, $\oli,\, i=1,\cdots,4$. They
were already determined in the original work \cite{Gasser}, but all of them 
are not accurate enough \cite{Ecker1}. The best values obtained so far
follows from matching the dispersion theoretic phenomenological
representation for $\pi\pi$ scattering amplitude to its two-loop evaluation 
in chiral perturbation theory \cite{Colangelo},
\[\ola=-0.4\pm 0.6, ~~~~~~ \olb=4.3\pm 0.1\,,~~~~~~~~~~~~\rm{(two\,loop)}\]
while the values relevant in the context of one-loop approximation are,
\[ \ola=-1.9\pm 0.2,~~~~~~ \olb=5.25\pm 0.04\,.~~~~~~~~~~~~\rm{(one\, loop)}\]
The difference in the two sets of values are attributed to the infrared
singularities that can be better dealt with in the two-loop matching than in
the case of one-loop. The original crude estimate of $\olc$ \cite{Gasser},
\[\olc=2.9\pm 2.4 \,,,~~~~~~~~~~~~~~~~~\rm{(one\,loop)}\]
has not been improved further. Finally the two-loop estimate of $\old$
\cite{Colangelo},         
\[\old=4.4\pm 0.2\,,~~~~~~~~~~~~~~\rm{(two\,loop)}\]
does not differ much from the original one-loop estimate \cite{Gasser},
\[\old=4.3\pm 0.9\,,~~~~~~~~~~~~~~~\rm{(one\,loop)}\]
as the infrared singularities are weakly present here. 

It should be noted here that
in our two-loop calculation of the pion pole in the axial-vector Green's
function at finite temperature, it is actually the scattering amplitude in
vacuum to one loop that enters its temperature dependent part. It is thus
appropriate to use the one-loop estimate of the coupling constants in the 
present context.

We now evaluate the pion pole parameters in two regions of temperature.
First consider the so-called high temperature limit, $T\gg\M$. To remain
within the domain of the low temperature expansion, this limit is
implemented not by letting $T$ increase, but instead by holding $T$ fixed
and sending $\M$ to zero. The value of $\M$ is determined by the quark
masses. Thus the high temperature limit is equivalent to the chiral limit of
$QCD$ theory.

The values of the relevant integrals in the chiral limit are given in the
Appendix C. The contributing terms are only $\eta^2\oj^2$ and the
combination, 
\be
\eta^2(l\,\oj_{00}+4\ok_{00})=\frac{T^4}{36F^4}(Z(T)+c),
\ee
where
\bea
& & Z(T)=\ln\frac{M}{T} +\frac{1}{10}(\ola+4\olb)\,,\nonumber\\
& & c=-\frac{7}{15} -\ln 2+1-I_1+I_2=0.30\,,\nonumber
\eea
as obtained from (4.13) and (C.1-2). Here $Z(T)$ has actually no logarithmic 
singularity in the chiral limit, as can be seen by shifting the 
renormalization scale of the coupling constants with Eq.(2.6) from $M$ to any 
other value $\mu$.

We thus get the results for the pion mass and decay constants to
two loops at finite temperature in the chiral limit as
\bea
& & \left.\frac{\M^2(T)}{\M^2}\right|_\chi = 1+\frac{T^2}{24F^2} -
\frac{T^4}{36F^4}\left(Z(T)+c+\frac{11}{32}\right) \nonumber \\
& &~~~~~~~~~~~~~= 1+\frac{T^2}{24F^2} -\frac{T^4}{36F^4}\ln\frac{\Lambda_M}{T}\\
& & \left.\frac{\F^t(T)}{\F}\right|_\chi = 1-\frac{T^2}{12F^2} +
\frac{T^4}{72F^4}\left(Z(T)+c+\frac{1}{4}\right) \nonumber \\
& &~~~~~~~~~~~~= 1-\frac{T^2}{12F^2} +\frac{T^4}{72F^4}\ln\frac{\Lambda_F}{T} \\
& & \left.\frac{\F^t(T)-\F^s(T)}{\F}\right|_\chi = \frac{T^4}{27F^4}
\left(Z(T)+c+\frac{1}{4}\right) \nonumber \\
& &~~~~~~~~~~~~~~~~~~~~~~~~= \frac{T^4}{27F^4}\ln\frac{\Lambda_\De}{T} 
\eea
where the logarithmic scales are 
\[\Lm_M=1.8 \,\rm{GeV},~~~~~~~~~ \Lm_F=\Lm_\De =1.6 \,\rm{GeV},\]
in agreement with Ref. \cite{Toublan}. Note that $\Lm_F$ is associated with
$\F^t(T)$ and not its square, as in this Ref. (Had we chosen the two-loop
coupling constants, we would get somewhat smaller values for the $\Lm$'s, 
namely, $\Lm_M=1.4$ GeV, $\Lm_F=\Lm_\De =1.3$ GeV.)
 
Next consider the low temperature limit, $T\ll \M$. We shall express all
quantities in terms of the dimensionless ratio
\[ \tau=\frac{T}{\M} \]
times possibly a power of temperature. 

The leading behaviour of all the pole parameters in the low temperature
region is given essentially by that of $\oj$, as seen from Eqs.(C.4-6). Thus we get
\bea
& &\left.\frac{\M^2(T)}{\M^2}\right|_\tau=1+\left(\frac{\M^2}{2\F^2}
-\frac{5g_1\M^4}{24\pi^2\F^4}\right) \oj|_\tau \\
& &\left.\frac{\F^t(T)}{\F}\right|_\tau=1-\left(\frac{\M^2}{\F^2}
-\frac{5g_2\M^4}{48\pi^2\F^4}\right) \oj|_\tau \\
& &\left.\frac{\F^t(T)-\F^s(T)}{\F}\right|_\tau=\frac{g_3\M^4}{12\pi^2\F^4}
\oj|_\tau
\eea
where
\bea
g_1 &=& \ola+2\olb-\frac{27}{40}\olc-\frac{3}{5}\old+\frac{9}{8},\nonumber\\
g_2 &=& \ola+2\olb-\frac{9}{10}\old+\frac{3}{10},\nonumber\\
g_3 &=& \ola+4\olb+\frac{4}{3},
\eea
and $\oj|_\tau$ is given by Eq.(C.4).

It is interesting to note that the two-loop contributions to the effective
parameters are always of opposite sign compared to that of one-loop, when it 
contributes. Figs. 5, 6 and 7 show the temperature dependence of
these parameters. Besides the pion interactions giving these results,
massive states are also expected to contribute. But as we show below, this
contribution turns out to be negligibly small up to $T\simeq 100$ MeV.   

\bef
\centerline{\psfig{figure=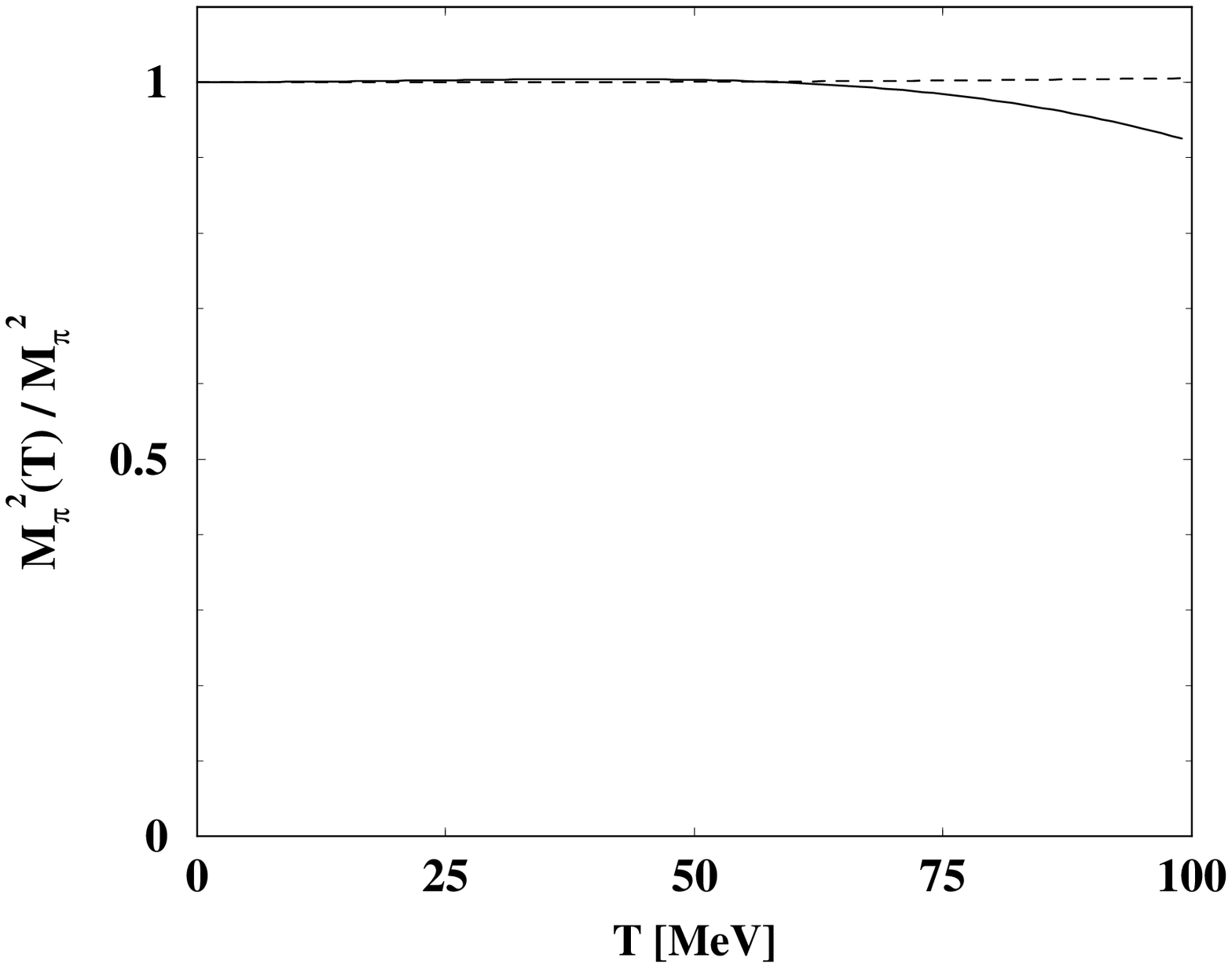,height=5cm,width=7cm}}
\caption{Thermal pion mass squared in chiral (continuous curve) and 
nonrelativistic (dashed one) limits.}
\eef
\bef
\centerline{\psfig{figure=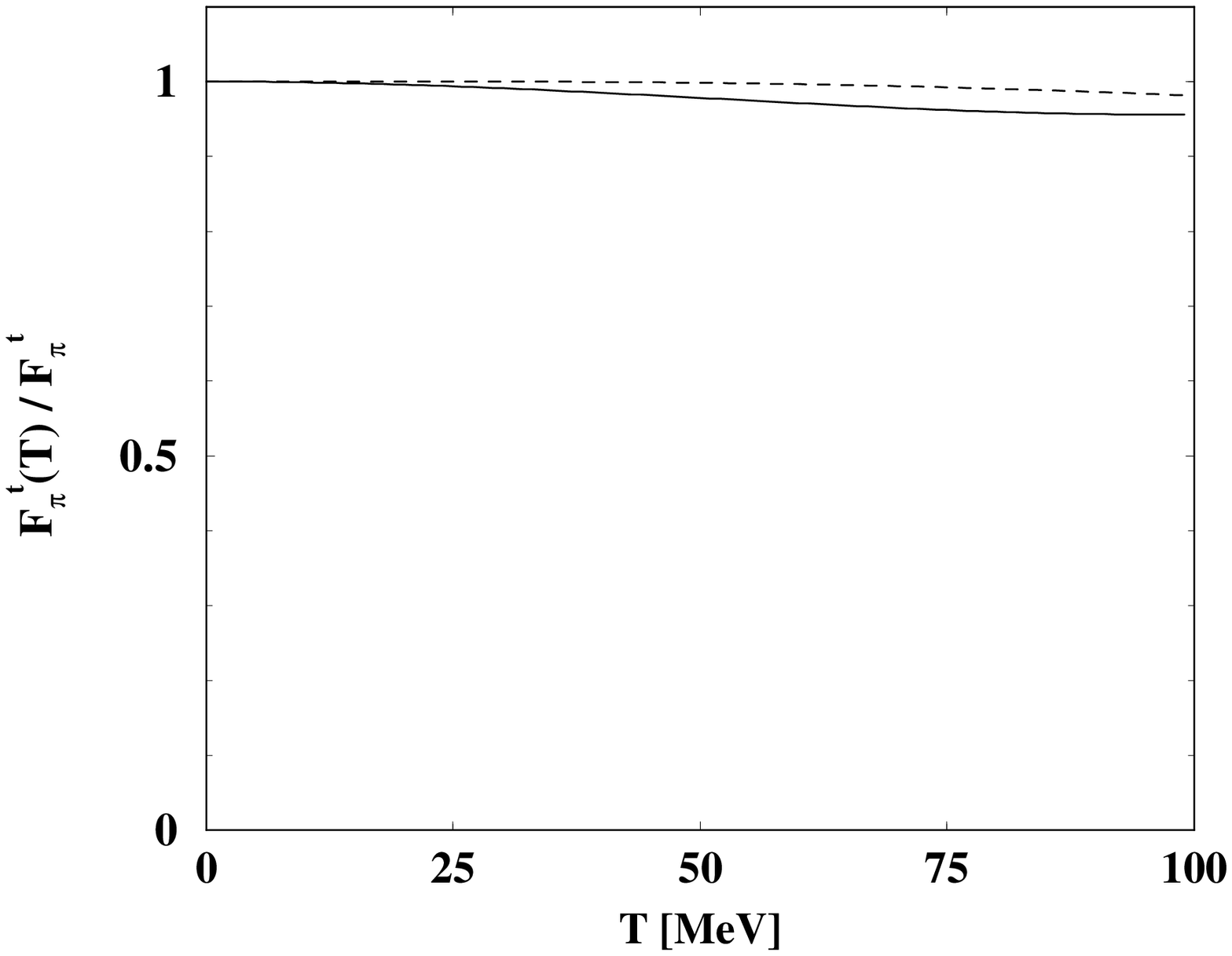,height=5cm,width=7cm}}
\caption{'Temporal' type of thermal pion decay constant in the two limits
as in Fig.5.} 
\eef

\bef
\centerline{\psfig{figure=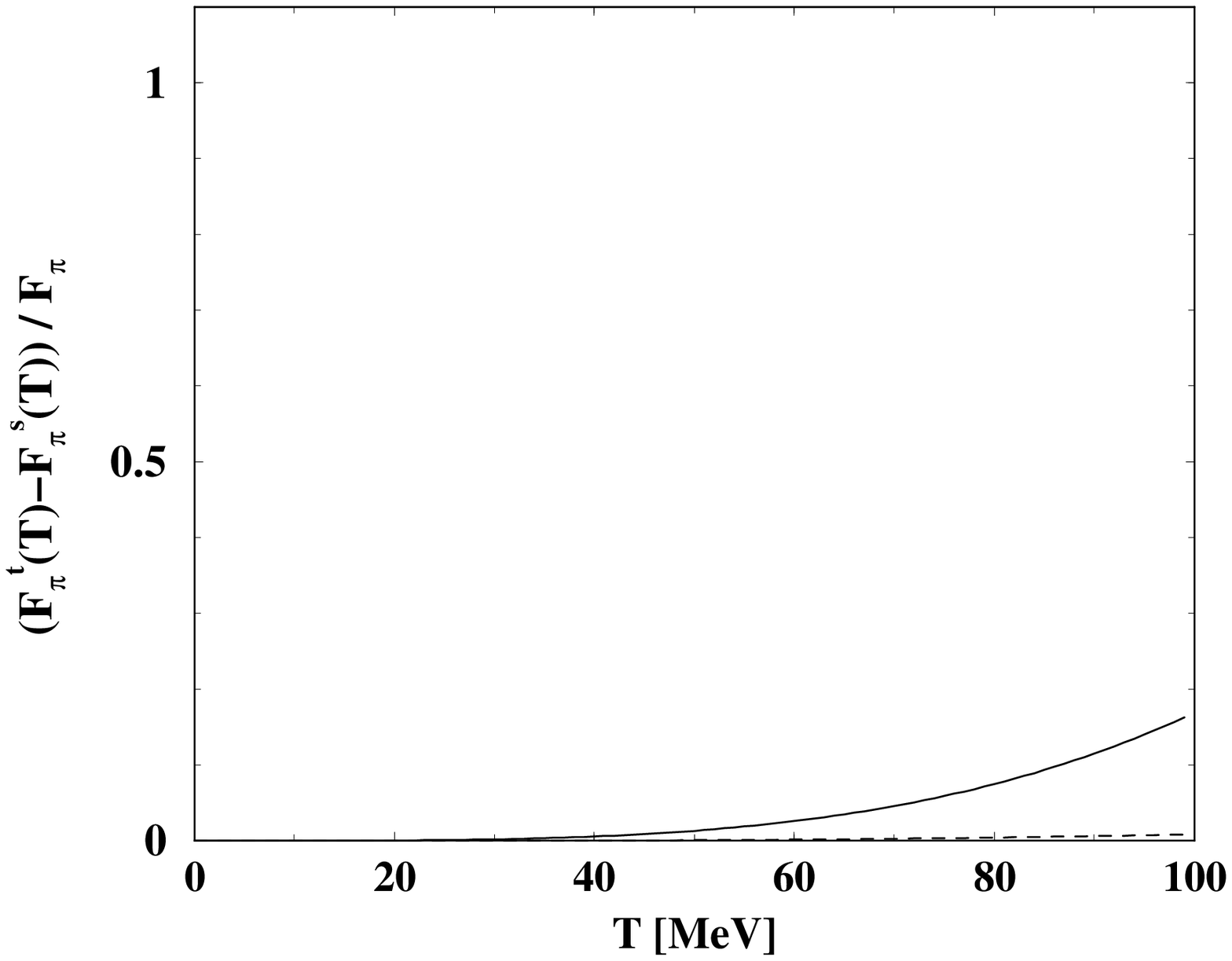,height=5cm,width=7cm}}
\caption{Difference of temporal and spatial types of pion decay constant 
in the two limits as in Figs. 5 and 6.}
\eef

\section{Massive states}
\setcounter{equation}{0}
\renewcommand{\theequation}{6.\arabic{equation}}

So far we considered only the pionic interactions in obtaining the pion pole
parameters. We now examine the leading contribution from the state(s) that
are massive in the chiral limit. Here the $\rho$ meson appears to be the most
important such state contributing through the graphs of Fig.8.

\bef
\centerline{\psfig{figure=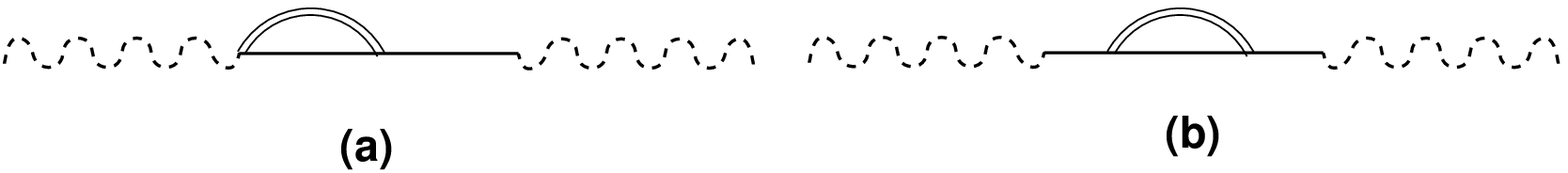,height=0.7cm,width=7.5cm}}
\caption{Graphs with $\rho$ propagator contributing to the pion pole}
\eef

The additional pieces of chiral Lagrangian needed for evaluation of the
graphs may be obtained from Ref.\cite{Ecker2,Ecker3}. Their construction follows 
from noting that the non-Goldstone fields transform only under the unbroken
isospin subgroup $SU(2)_V$ with its group parameters dependent on the Goldstone 
fields. Accordingly one has to replace the ordinary derivative on the $\rho$ meson 
field $\rho_\mu (x)$ by the covariant one,
\[\nabla_\mu\rho_\nu=\dmd\rho_\nu+[\Gm_\mu,\rho_\nu]. \]
Here the connection $\Gm_\mu$ is given by
\[\Gm_\mu=\frac{1}{2}u^\dag (\dmd-ia_\mu)u+\frac{1}{2}u(\dmd+ia_\mu)u^\dag,
~~~~~~u^2=U,\]
where, as in Eq.(2.1), we have set the external vector field to zero. One  
need also construct appropriate variables from $D_\mu U$ and $F_{R,L}^{\mn}$, 
the right- and the left-handed field strengths built out of the external 
vector and the axial-vector potentials, that transform only under $SU(2)_V$, 
namely
\[u_\mu=iu^\dag D_\mu Uu^\dag = u_\mu^\dag\]
and 
\[f^{\mn}_{\pm}=\pm u^\dag F_R^{\mn}u+u F_L^{\mn}u^\dag. \]

In terms of the above variables, one can write the required pieces of the
chirally invariant Lagrangian as \cite{Ecker2,Ecker3},
\be
{\cal L}_{int}(\rho)=\frac{1}{2\sqrt{2}m_\rho}(F_\rho \la
\rho_{\mn}f_+^{\mn} \ra +iG_\rho\la \rho_{\mn}[u^\mu , u^\nu]\ra),
\ee
where $\rho_{\mn}=\nabla_\mu\rho_\nu-\nabla_\nu\rho_\mu$. From the decay
rates $\Gm (\rho^0 \rw e^+e^-)$ and $\Gm (\rho \rw 2\pi)$, one gets
respectively $F_\rho=154$ MeV and $G_\rho=69$ MeV \cite{Ecker2}.

It is now simple to calculate the vacuum amplitudes of graphs in Fig.~8,
which may then be converted to their thermal counterparts, following
discussions in Appendix A. Now the $\rho$ particles are rather scarce in the
medium for temperatures up to about $100$ MeV \cite{Gerber}. Accordingly 
we drop the $T$-dependent term in the $\rho$ propagator. Further, as
the momenta involved are small compared to $m_\rho$, we keep only the
leading term in the expansion of the vacuum $\rho$ propagator in inverse 
powers of $m_\rho^2$. The amplitudes thus simplified can be expressed through
$J^\beta$ and $J_{\mn}^\beta$ only. 

The self-energy graph contributes a $T$-dependent amplitude,
\be
F^2q_\mu q_\nu 24\frac{G_\rho^2M^2}{m_\rho^4}\eta^2(\oj-\oj_{00})\left(
i\De (q) +\frac{2}{3}\De^2(q)\right). 
\ee
Clearly it vanishes in the chiral (high $T$) limit. Also its leading
contribution vanishes in the region where pions move nonrelativistically
(low $T$ region). In the same way, the vertex graph contributes a $T$-dependent 
part,
\be
F^2 8\frac{F_\rho G_\rho M^2_\pi}{m_\rho^4} \eta^2 \{-2\oj_{00} q_\mu q_\nu
+(q_\mu\oj_{\nu\lm}+\oj_\nu \oj_{\mu\lm})q^\lm\}i\De (q),
\ee
which again vanishes in the chiral limit, but is finite in the low temperature 
region. Surprisingly, it contributes only to $\F^s(t)$,
\be
\left.\frac{\F^s(T)}{\F}\right|_{(\rho)\tau}=-8\frac{\M^4}{\F^4}
\frac{F_\rho G_\rho\M^2}{m_\rho^4}\oj~|_\tau
\ee
This contribution, however, is only about $3\%$ of the pionic contribution
in Eq.(5.7).

\section{Discussion}

Though the propagators, vertices and self-energies assume the
form of $2\times 2$ matrices in the real time field theory at finite
temperature, each of them is essentially given by a single analytic 
function, as can be seen from an appropriate factorization of these matrices. 
The same is true of the ensemble average of (the $T$-product of) any two 
operators. Here we show that one can find its thermal $2\times 2$ matrix
amplitude directly from the vacuum amplitude and take advantage of this
factorization to get the analytic (single component) thermal amplitude.
Thus compared to the commonly followed practice of considering the
11-element of the thermal matrix, the use of matrices not only simplifies
and frees the calculation from ill-defined quantities at intermediate
steps, but also yields directly the amplitude with proper analytic
properties, not possessed by the 11-element.

In this work we use this matrix method to calculate the thermal pion pole
term in the axial-vector two-point function in the framework of chiral
perturbation theory. From the analytic amplitude we derive the effective
mass and the decay constants of the pion at finite temperature. These are
evaluated in two limits, the chiral and the nonrelativistic. The two 
evaluations agree rather closely up to about $T\simeq 100$ MeV. We also 
examine the contribution of the massive states ($\rho$) to the effective 
parameters and find it to be negligible.  

Finally we compare our work with that of Toublan \cite{Toublan},
whom we follow at a number of points. He obtains the thermal amplitude in a
somewhat intuitive manner, while we formulate rules to write the
matrix amplitude, which leads immediately to the analytic thermal amplitude.
These rules, in effect, justify his way of writing the thermal amplitude, as
far as its real part is concerned. Our results (4.18-20) for the effective
mass and decay constants of the pion agree with his, except for the 
the coefficient $l'''$ in the equation for $\M^2(T)$ \cite{Comment}. But  
their chiral limits (5.2-4) agree completely with his, as the term 
with this coefficient does not contribute in this limit. We also find the 
effective parameters in the nonrelativistic limit, in which this term 
does contribute. In his discussion of the effect of massive states on the
pion parameters, he considers only the real states in the heat bath and
concludes it to be negligible up to $T \simeq 100$ MeV. We also consider the 
effect of the virtual massive states in vacuum and find it too to be negligible. 

\section*{Appendix A}
\setcounter{equation}{0}
\renewcommand{\theequation}{A.\arabic{equation}}

The feature of the real time thermal field theory that distinguishes it from
the vacuum theory is the time contour in their generating functionals. While
it is the infinite real line for the vacuum theory, it must be augmented with a
return path for the thermal theory. One example of such a path that we
shall use is shown in Fig.~5. The return path may be folded onto the onward
path, generating fields with a displaced time argument. Thus compared with
the vacuum generating functional,
\be
\la 0|T\exp [i\int d^3x\int_{-\infty}^{\infty} dt
\{{\cl}_{int}(\phi)+j(x)\phi(x)+j^\mu(x)A_\mu(x)\}_{in}]|0\ra\,,
\ee
the thermal one is given by
\be
{\rm Tr} \left[ \rho T\exp  [i\int d^3x\int_{-\infty}^{\infty} dt
\{{\cl}_{int}(\phi_1)-{\cl}_{int}(\phi_2)
+\sum_{n=1,2}(j_n(x)\phi_n(x)+j^\mu_n(x)A_{\mu\,n}(x))\}_{in} ]\right]\,,
\ee
both written in the interaction representation in terms of the in-fields,
denoted by the subscript 'in', which we shall omit below. 
The fields $\phi_1(x)$ and $\phi_2(x)$ have their time arguments on the 
segments $C_1$ and $C_2$ respectively: $\phi_1(x)=  \phi (\vec{x},t)$ is the
'physical field' and $\phi_2(x)= \phi (\vec{x}, t-i\bet/2)$ is the 'ghost'
field. The $-$ sign before ${\cl}(\phi_2)$ is forced upon us by the theory,
but the signs before the $n=2$ terms with the external fields are at our
disposal, which we choose to be positive. The pieces $C_3$ and $C_4$ of the 
complex time contour are of no consequence and have been dropped.

\bef
\centerline{\psfig{figure=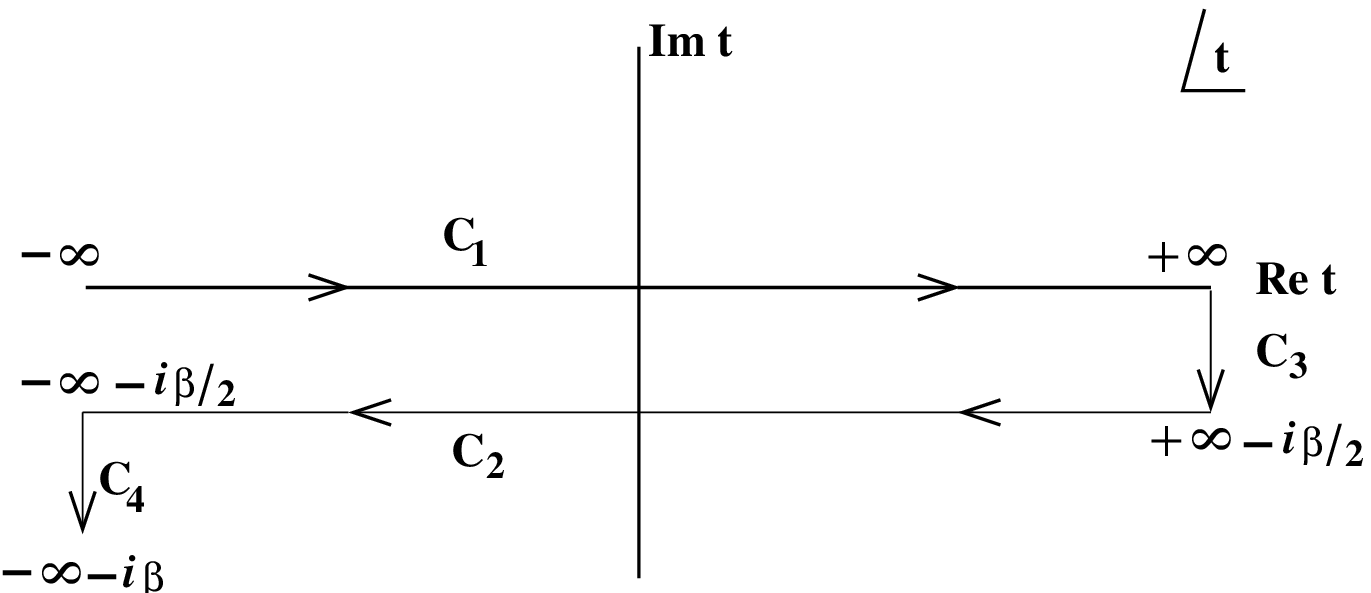,height=3.5cm,width=8.5cm}}
\caption{The complex time contour of real time thermal field theory}
\eef
  
The two sets of fields make any thermal two-point function a $2\times 2$
matrix. In particular, the free pion propagator is
\be
\bde^{ab}(x-y) = \de^{ab}\bde (x-y)= \left( \begin{array}{cc}
\la T \phi_1^a(x)\phi_1^b(y) \ra &
\la \phi_2^b(y) \phi_1^a(x) \ra \\
\la \phi_2^a(x) \phi_1^b(y) \ra &
\la \tilde{T} \phi_2^a(x)\phi_2^b(y) \ra
\end{array} \right)\,,
\ee
where $\tilde{T}$ denotes anti-time ordering, can be evaluated directly 
in momentum space as,
\be
\bde (q)=\left( \begin{array}{cc}
\De (q)+2\pi n(|q_0|)\de(q^2-M^2) &
2\pi n(|q_0|)\de(q^2-M^2)e^{\bet |q_0|/2}\\
2\pi n(|q_0|)\de(q^2-M^2)e^{\bet |q_0|/2} &
\De^* (q)+2\pi n(|q_0|)\de(q^2-M^2)
\end{array} \right)
\ee
where $n(|q_0|)=(e^{\bet |q_0|}-1)^{-1}$ is the pion distribution function and 
$\De (q)=i/(q^2-M^2+i\ep)$ is the pion propagator in vacuum. (A boldface
letter will always indicate a $2\times 2$ matrix.) 

Given the Feynman amplitude for any two-point function in vacuum, it is simple
to write the corresponding thermal matrix amplitude. The correspondence may
be found by comparing the Wick contractions for graphs in the two
cases, with particular attention to the $-$ sign before the 'ghost' Lagrangian.
As an example, consider the graph (d) of Fig.~1, for which we show this
correspondence in detail. Its vacuum amplitude in coordinate space is 
obtained from
\be
-F^2i\int d^4z \la 0| T\dmd\phi^a(x)\dnd \phi^b(y) {\cl}_{int} (\phi(z))|0\ra, 
\ee
where ${\cl}_{int}$ is a piece in ${\cl}^{(2)}$,
\be
{\cl}_{int}(\phi)=-\frac{1}{6F^2} \left\{ \p\cdot\p\dmd\p\cdot\dmu\p
-\p\cdot\dmd\p\p\cdot\dmu\p -\frac{M^2}{4} (\p\cdot\p)^2 \right\}\,,
\ee
the field $\p$ denoting the pion iso-vector triplet $(\phi^1,
\phi^2,\phi^3)$. It gives the amplitude in momentum space,
\be
T_{\mn}^{(1d)}(q)=q_\mu q_\nu J(M) M^2\left \{\frac{2}{3}i\De(q) +
\frac{1}{2}M^2 \De^2(q)\right\}\,,
\ee
where $J(M)$ is defined by Eq.(3.3). The single propagator in this expression 
arises from the cancellation,
\be
(q^2-M^2)\De^2(q)=i\De(q)\,.
\ee
We now identify the contractions in (A.5) that produce this result. To focus on the 
contractions, we omit the derivatives and isospin indices on the pion fields and 
write schematically a term of the matrix element (A.5) as
\bea
& &\la 0|T\phi(x)\phi(y)\phi^4(z)|0\ra \nonumber \\
& \sim & J(M) \De(x-z)\De(z-y) \nonumber \\
& \sim & J(M)\De^2(q)\,,
\eea 
in momentum space. To get the corresponding thermal matrix amplitude, 
consider its $ij$-th element,
\be
-F^2i\int d^4z \la  T\dmd\phi_i^a(x)\dnd \phi_j^b(y) 
\{{\cl}_{int}(\phi_1(z))-{\cl}_{int}(\phi_2(z))\} \ra.
\ee
Again we write schematically a term of this matrix element and contract its
fields as,
\bea
& &\la T\phi_i(x)\phi_j(y)\{\phi_1^4(z)-\phi_2^4(z)\}\ra \nonumber \\
& \sim & J^\beta (M) \la T\phi_i(x)\phi_j(y)\{\phi_1^2(z)-\phi_2^2(z)\}\ra 
\nonumber \\
& \sim & J^\beta (M)\{\bde_{i1}(x-z)\bde_{1j}(z-y)-\bde_{i2}(x-z)\bde_{2j}(z-y)\} 
\nonumber \\
& \sim & J^\beta (M)(\bde(q)\bt\bde(q))_{ij}\,,
\eea
in momentum space, where we use in the second line the fact that the
contractions of two $\phi_1$'s and two $\phi_2$'s at the same point yield 
the same result,
\be
J^\beta (M)=\frac{1}{M^2}\int \frac{d^4k}{(2\pi)^4}\bde (q)_{11\,or\,22}
\ee
and the matrix $\bt$ is 
\be
\bt= \left( \begin{array}{cc}
1 & 0 \\
0 & -1
\end{array} \right)\,.
\ee
Note that a cancellation similar to Eq.(A.8) for the vacuum case works also
here,
\be
(q^2-M^2)\bde (q)\bt\bde (q)=i\bde(q)\,.
\ee
Comparing the contractions (A.9) and (A.11), we see that the thermal matrix 
amplitude of graph(1d) can be obtained from the vacuum amplitude (A.7) simply by 
replacing $J$ with $J^\beta$, $\De$ with $\bde$ and $\De^2$ by $\bde\bt\bde$ in it.

A little reflection on other graphs of Figs.~1-3 will convince us that the
thermal amplitudes of all these graphs may be obtained from their vacuum
amplitudes by the replacements just stated above, together with 
$J^{\prime}$ and $J_{\mn}$ by $J^{\prime\, \beta}$ and $J^\beta_{\mn}$
respectively, where
\bea
J^{\prime\,\beta}(M)&=&i\int\frac{d^4k}{(2\pi)^4}(\bde(k)\bt\bde(k))_{11\,or\,22}
\nonumber \\ 
&=&-\frac{\partial}{\partial M^2}(M^2J^\beta)\,,
\eea
and
\be
J^\beta_{\mn}(M)=\frac{1}{M^4}\int\frac{d^4k}{(2\pi)^4}\,k_\mu k_\nu
\bde(k)_{11\,or\,22}
\ee
In Eq.(A.15) we use the so-called mass-derivative
formula for the matrix propagator \cite{Fujimoto},
\be
(\bde\bt)^2=i\frac{\partial}{\partial M^2} (\bde\bt)\,,
\ee
which is the thermal extension of the trivial relation 
$\De^2(q)=i\partial \De/\partial M^2$ for the vacuum propagator.

So long we constructed thermal amplitudes with the matrix propagator only,
inserting $\bt$ explicitly to account for the $(-)$ sign from the 'ghost'
Lagrangian. We now introduce two kinds of parts of graphs namely, 
the self-energy and the (two-point) vertices, where it is convenient to 
include the effect of the associated $(-)$ sign(s) in their definitions. 
Thus writing the $ij$-th component of the matrix amplitude of Fig.4b,
again schematically and contracting the fields, we get
\bea
& &\la T\phi_i(x)\{\phi^4_1(u)-\phi^4_2(u)\}
\{\phi^4_1(v)-\phi^4_2(v)\}\phi_j(y)\ra \nonumber \\
& & \sim (\bde (x-u)\bsg (u-v)\bde (v-y))_{ij}\,,
\eea
absorbing the $(-)$ signs in the definition of $\bsg$,
\be
\bsg= \left( \begin{array}{cc}
s_{11} & -s_{12} \\
-s_{21} & s_{22}
\end{array} \right),~~~~~s_{ij}=({\bde}_{ij})^3
\ee
Likewise, for the graph of Fig.~4a we write
\be
\la T \phi_i^3(x)\{\phi_1^4(u)-\phi_2^4(u)\}\phi_j(y)\ra \sim
(\bgm^{(1)}\bde)_{ij},~~~~\bgm^{(1)}= \left( \begin{array}{cc}
s_{11} & -s_{12} \\
s_{21} &-s_{22}
\end{array} \right),
\ee
and
\be
\la T \phi_i(x)\{\phi_1^4(u)-\phi_2^4(u)\}\phi_j^3(y)\ra \sim
(\bde \bgm^{(2)})_{ij},~~~~\bgm^{(2)}= \left( \begin{array}{cc}
s_{11} & s_{12} \\
-s_{21} &-s_{22}
\end{array} \right)\,.
\ee

The matrix amplitudes are greatly simplified by factoring out matrices
involving only the pion distribution function. Thus the free propagator
given by Eq.(A.4) can be factored as 
\be
\bde (q)=\U(q)\left( \begin{array}{cc}
\De(q) & 0\\
0 & \De^{\ast}(q)
\end{array} \right)\U(q)~,~~~
\U(q)=\left( \begin{array}{cc}
\sqrt{1+n} & \sqrt{n} \\
\sqrt{n} & \sqrt{1+n}
\end{array} \right)~.
\ee
Also the full propagator $\bde'$ and the two-point function 
$\T_{\mn}$ admit similar factorizations,
\be
\bde^{\prime} (q)=\U(q)\left( \begin{array}{cc}
\De^{\prime\,\beta}(q) & 0\\
0 & \bde^{\prime\,\beta\,\ast}(q)
\end{array} \right)\U(q)~,~~~
\T_{\mn} (q)=\U(q)\left( \begin{array}{cc}
T_{\mn}^\beta (q) & 0\\
0 & -T_{\mn}^{\beta\,\ast}(q)
\end{array} \right)\U(q)~,
\ee
as is suggested by evaluation of our graphs. More rigorously, these follow
from their spectral representations.

To derive a similar factorization of the self-energy part $\bsg$, we look at
the Dyson-Schwinger equation for the full propagator,
\be
\bde\pr=\bde+\bde(-i\bsg)\bde\pr \,.
\ee
Inserting the factorizations for $\bde'$ and $\bde$, we infer that 
$\bsg$ must have the factorized form \cite{Kobes},
\be
\bsg (q)=\U^{-1}\left( \begin{array}{cc}
\Sg^{\beta}(q) & 0\\
0 & -\Sg^{\beta\,\ast}(q)
\end{array} \right)\U^{-1}~.
\ee
It immediately follows that 
\be
\bsg_{22}=-\bsg_{11}^\ast,~~~~ \bsg_{21}=\bsg_{12}\,.
\ee
Further we can get the function $\Sg^{\beta}$ entirely from $\bsg_{11}$,
\be
Re\,\Sg^{\beta}=Re\, \bsg_{11},~~~~ Im\,\Sg^\beta =\frac{1}{1+2n}
Im\,\bsg_{11}\,.
\ee
In the same way the relations,
\be
\T\sim -i\bgm^{(1)}\bde \sim \bde (-i\bgm^{(2)})
\ee
give us the factorizations,
\be
\bgm^{(1)}=\U\left( \begin{array}{ll}
\Gm^\bet & 0\\
0 & \Gm^{\bet\,\ast}
\end{array} \right)\U^{-1}~,~~~~~
\bgm^{(2)}=\U^{-1}\left( \begin{array}{ll}
\Gm^\bet & 0\\
0 & \Gm^{\bet\,\ast}
\end{array} \right)\U
\ee
We see that $\bgm^{(1)}$ and $\bgm^{(2)}$ differ by a $-$ sign in the 
off-diagonal elements, which is of no consequence to us. We thus omit the
superscripts to write the relations given by Eq.(A.29) as,
\be
\bgm_{22}=\bgm_{11}^{\ast},~~ \bgm_{21}=\bgm_{12}.
\ee
Again the $11$ element determines the function $\Gm^{\beta}$ completely,
\be
Re\Gm^\bet=Re\bgm_{11},~~~Im\Gm^\bet=\frac{1}{1+2n}\, Im\bgm_{11}
\ee

\section*{Appendix B}
\setcounter{equation}{0}
\renewcommand{\theequation}{B.\arabic{equation}}

The $\bet$-dependent, divergent and finite parts of $K^{\bet}(q)$ and
$K^{\bet}_{\mn}$ have been obtained in Ref.\cite{Toublan}, which we reproduce 
here for completeness. The divergent parts reside only in terms linear in the
distribution function, where we need the vacuum integrals,
\be
L(p)=i\int\frac{d^4k}{(2\pi)^4}\De(k)\De(p-k)\nonumber\\
=-2\lm-\frac{1}{16\pi^2} +R(p)\,,
\ee
and
\bea
L_{\mn}(p)&=&\frac{i}{M^2}
\int\frac{d^4k}{(2\pi)^4}k_\mu k_\nu\De(k)\De(p-k)\nonumber\\
&=& \frac{\lm }{6 M^2}\{ (p^2 -6 M^2)g_{\mn}-4p_\mu p_\nu\}\nonumber\\
&& -\frac{1}{2(24\pi M)^2}[20p_\mu p_\nu-2p^2g_{\mn}
+\{4p_\mu p_\nu -(p^2-4M^2)g_{\mn}-4M^2p_\mu p_\nu /p^2\}R(p)]
\eea
with
\be
R(p)=-\frac{1}{16\pi^2}\int_0^1 dx \ln (1-p^2x(1-x)/M^2)
\ee
Then the terms linear in $n$ are given by
\bea
K^\bet(q)|_n&=&\frac{3}{M^2}\int\frac{d^4k}{(2\pi)^3}\de(k^2-M^2)n(|k_0|)L(q-k)~,
\nonumber \\
&=& -6\lm\oj-\frac{3}{16\pi^2}\oj+
\frac{3}{M^2}\int\frac{d^4k}{(2\pi)^3}\de(k^2-M^2)n(|k_0|)R(q-k)
\eea
and
\bea
K^\bet_{\mn}(q)|_n &=&\frac{1}{M^4}\int\frac{d^4k}{(2\pi)^3}k_\mu k_\nu
\de(k^2-M^2)n(|k_0|)L(q-k)
+\frac{2}{M^2}\int\frac{d^4k}{(2\pi)^3}k_\mu k_\nu\de(k^2-M^2)n(|k_0|)
L_{\mn}(q-k)\nonumber \\
&=& -\frac{\lm}{3}[10\oj_{\mn}+\{5g_{\mn}+(4q_\mu q_\nu-q^2g_{\mn})/M^2\}\oj]
\nonumber \\
&&+\frac{1}{2(12\pi M)^2}[\{(q^2+M^2)g_{\mn}-10q_\mu q_\nu\}\oj-28M^2
\oj_{\mn}] \nonumber \\
&&+\frac{1}{6M^4}\int\frac{d^4k}{(2\pi)^3}\de (k^2-M^2)n(|k_0|)R(q-k)\cdot
\nonumber \\
&& [4q_\mu q_\nu -4(q_\mu k_\nu+q_\nu k_\mu)+10k_\mu k_\nu 
-g_{\mn}(q^2-3M^2-2q\cdot k)-4M^2(q-k)_\mu (q-k)_\nu/(q-k)^2 ]
\eea

Next consider the terms quadratic in $n$. Introducing the integrals,
\be
Q(p)=\int\frac{d^4k}{(2\pi)^3}\frac{\de (k^2-M^2)n(k)}{(p-k)^2-M^2}\,,~~~~
Q_\mu(p)=\int\frac{d^4k}{(2\pi)^3}k_\mu\frac{\de (k^2-M^2)n(k)}{(p-k)^2-M^2}\,, 
\ee
we can write them as
\be
K^\bet (q)|_{n^2}=-\frac{3}{M^2}\int\frac{d^4k}{(2\pi)^3}\de (k^2-M^2)n(|k_0|)
Q(q-k)
\ee
and
\bea
K_{\mn}^\bet (q)|_{n^2}=-\frac{1}{M^4}\int\frac{d^4k}{(2\pi)^3}
\de (k^2-M^2)n(|k_0|)&&[\{q_\mu q_\nu +4k_\mu k_\nu -2(q_\mu k_\nu +q_\nu k_\mu)\}
Q(q-k) \nonumber\\
&& + k_\mu Q_\nu (q-k)+k_\nu Q_\mu (q-k)]
\eea
These terms as well as the last terms in Eqs.(B.3) and (B.4) belong to 
$\ok (q)$ and $\ok_{\mn} (q)$ in the notation of Eqs. (4.7) and (4.8).

Note that $Q_i$ and $Q_0$ are not independent, but are related by
\be
Q_i(p)=\frac{p_i}{2|\vp|^2}\{2p_0Q_0(p)-p^2Q(p)+\oj\}\,.
\ee
Using this relation one can relate $\ok_{ij}$ and $\ok_{00}$ at the pole,
$q_0=M$ with $\vq=0$,
\be
\ok_{ij}=\frac{\de_{ij}}{3}(\ok_{00}-\ok+\oj^2)
\ee

Being interested in the real parts, we shall not consider terms cubic in
$n$, which are imaginary.

\section*{Appendix C}
\setcounter{equation}{0}
\renewcommand{\theequation}{C.\arabic{equation}}

Here we write the chiral and the nonrelativistic limits of the integrals
occurring in expressions for the effective pion parameters. The chiral limits  
were obtained in Ref.\cite{Toublan}\footnote{In Ref.\cite{Toublan} the
integral $I_1$ is evaluated by expressing it in terms of derivatives of
Zeta and Gamma functions. But since the integral $I_2$ has to be evaluated
numerically anyway, we can do the same for $I_1$ also and get identical
result.},
\be
\left.\eta\oj\right|_\chi=\frac{T^2}{12F^2}-\frac{MT}{4\pi F^2}\,,~~~~~~~~
\left.\eta^2\oj_{00}\right|_\chi=\frac{\pi^2T^4}{30F^4}
\ee
\be
\left.\eta^2\ok_{00}\right|_\chi=\frac{T^4}{144F^4}\left( \ln\frac{M}{T}
-\ln 2+1- I_1 + I_2
\right)
\ee
where
\bea
& & I_1=\frac{15}{2\pi^4}\int_0^{\infty}\frac{dx\,x^3\ln x}{e^x-1} =0.60\,,
\nonumber\\
& & I_2=\frac{18}{\pi^4}\int_0^{\infty}\frac{dx\,x^3}{e^x-1}\int_0^1
\frac{d\al}{e^{\al x}-1}\left\{ (1+\al^2)\ln\frac{1+\al}{1-\al}
+\al\ln\frac{1-\al^2}{\al^2 x^2}\right\}=1.05 
\eea
The integrals for $\eta^2\ok|_\chi$ and $\eta^2\frac{\del\ok}{\del q_0}|_\chi$ vanish,
while those for $\eta^2\frac{\del\ok_{00}}{\del q_0}|_\chi$ and $\eta^2 C|_\chi$ 
are finite in the chiral limit. (Actually the terms linear and quadratic in $n$ in 
each of the later two quantities have singular pieces separately in the this
limit, but they cancel out in their respective sums.)

Next we calculate the integrals in the low temperature region, where
$\tau\equiv T/\M\ll 1$. Keeping only the leading terms, we get
\bea
& & \left.\oj\,\right|_\tau =\left.\oj_{00}\right|_\tau
= \left(\frac{\tau}{2\pi}\right)^{3/2} e^{-\frac{1}{\tau}}\\
& &\frac{1}{3} \left.\ok\,\right|_\tau  =\frac{1}{3}\left.\ok_{00}\right|_\tau
=\frac{1}{2}\M \left.\frac{\del \ok_{00}}{\del q_0}\right|_\tau=\M C\,|_\tau
=\frac{1}{16\pi^2}\oj\,|_\tau \\
& &\M\left.\frac{\del\ok}{\del q_0}\right|_\tau= 
O(\tau^{5/2} e^{-\frac{1}{\tau}}).
\eea

\section*{Acknowledgement}

One of us (S.M.) acknowledges earlier support of CSIR, Government of India.
He also thanks Prof. H. Leutwyler for a correspondence.

\end{document}